\documentclass[letter, 11pt]{article}

\usepackage{geometry}

\usepackage{amsmath, amsthm, amssymb, verbatim}
\usepackage{epsfig}
\usepackage{graphics}

\newtheorem{proposition}{Proposition}

\newtheorem{theorem}{Theorem}
\newtheorem{lemma}{Lemma}

\newcommand{\StackMST}{\textsc{StackMST} }
\newcommand{\StackMSTnospace}{\textsc{StackMST}}
\newcommand{\scover}{\textsc{SetCover} }
\newcommand{\scovernospace}{\textsc{SetCover}}
\newcommand{\vcover}{\textsc{VertexCover} }
\newcommand{\ds}{\displaystyle}
\newcommand{\OPT}{\mathrm{OPT}}
\newcommand{\OPTVC}{\mathrm{OPT}_{\mathrm{VC}}}
\newcommand{\IP}{\mathrm{IP}}
\newcommand{\LP}{\mathrm{LP}}

\begin{document}

\title{The Stackelberg Minimum Spanning Tree Game%
\thanks{A preliminary version of this article appeared in the Proceedings of 
the 10th Workshop on Algorithms and Data Structures (WADS 2007), see~\cite{CDFJLNW07}.
This work was partially supported by the {\em Actions de Recherche Concert\'ees (ARC)\,} 
fund of the {\em Communaut\'e fran\c{c}aise de Belgique}.}}

\author{
Jean Cardinal\thanks{Universit\'e Libre de Bruxelles, D\'epartement d'Informatique, c.p.~212, B-1050 Brussels, Belgium, jcardin@ulb.ac.be.}
\and 
Erik D. Demaine\thanks{MIT Computer Science and Artificial Intelligence Laboratory, Cambridge, MA 02139, USA, edemaine@mit.edu.}
\and
Samuel Fiorini\thanks{Universit\'e Libre de Bruxelles, D\'epartement de Math\'ematique, c.p.~216,  B-1050 Brussels, Belgium,  sfiorini@ulb.ac.be.}
\and 
Gwena\"el Joret\thanks{Universit\'e Libre de Bruxelles, D\'epartement d'Informatique, c.p.~212,  B-1050 Brussels, Belgium, gjoret@ulb.ac.be. G. Joret is a 
Postdoctoral Researcher of the Fonds National de la Recherche Scientifique (F.R.S.--FNRS).}
\and
Stefan Langerman\thanks{Universit\'e Libre de Bruxelles, D\'epartement d'Informatique, c.p.~212, B-1050 Brussels, Belgium, slanger@ulb.ac.be. S. Langerman is a Research Associate of the Fonds National de la Recherche Scientifique (F.R.S.--FNRS).}
\and
Ilan Newman\thanks{Department of Computer Science, University of Haifa, Haifa 31905, Israel, ilan@cs.haifa.ac.il.} 
\and 
Oren Weimann\thanks{MIT Computer Science and Artificial Intelligence Laboratory, Cambridge, MA 02139, USA, oweimann@mit.edu.}
}

\date{}

\pagestyle{plain} \maketitle

\begin{abstract}

We consider a one-round two-player network pricing game,
the {\em Stackelberg Minimum Spanning Tree\/} game or \StackMSTnospace.

The game is played on a graph (representing a network),
whose edges are colored either red or blue,
and where the red edges have a given fixed cost
(representing the competitor's prices).
The first player chooses an assignment of prices to the blue edges,
and the second player then buys the cheapest possible minimum spanning tree,
using any combination of red and blue edges. The goal of the first player is to 
maximize the total price of purchased blue edges. 
This game is the minimum spanning tree analog of the well-studied
Stackelberg shortest-path game.

{}\quad
We analyze the complexity and approximability of the first player's
best strategy in \StackMSTnospace.
In particular, we prove that the problem is APX-hard even if there are
only two different red costs, and give an approximation algorithm whose
approximation ratio is at most $\min \{k,1+\ln b,1+\ln W\}$,
where $k$ is the number of distinct red costs, $b$ is the number of
blue edges, and $W$ is the maximum ratio between red costs.
We also give a natural integer linear programming formulation of the problem,
and show that the integrality gap of the fractional relaxation
asymptotically matches the approximation guarantee of our algorithm.
\end{abstract}

\section{Introduction}

Suppose that you work for a networking company that owns many
point-to-point connections between several locations,
and your job is to sell these connections.  A customer
wants to construct a network connecting all pairs of locations in the
form of a spanning tree. The customer can buy connections that you are
selling, but can also buy connections offered by your competitors.
The customer will always buy the cheapest possible spanning tree.
Your company has researched the price of each connection
offered by the competitors.
The problem considered in this paper is how to set the price of each
of your connections in order to maximize your revenue, that is,
the sum of the prices of the connections that the customer buys from you.

This problem can be cast as a {\em Stackelberg game}, a type of two-player
game introduced by the German economist Heinrich Freiherr von
Stackelberg~\cite{Stack34}. In a Stackelberg game, there are two players:
the {\em leader\,} moves first, then the {\em follower\,} moves,
and then the game is over.
The follower thus optimizes its own objective function, knowing the
leader's move. The leader has to optimize its own objective function
by anticipating the optimal response of the follower.
In the scenario depicted in the preceding paragraph,
you were the leader and the customer was the follower:
you decided how to set the prices for the connections that you own,
and then the customer selected a minimum spanning tree.
In this situation, there is an obvious tradeoff: the
leader should not put too high price on the connections---otherwise
the customer will not buy them---but on the other hand the leader needs to
put sufficiently high prices to optimize revenue.

Formally, the problem we consider is defined as follows.
We are given an undirected graph\footnote{All graphs in this paper are finite 
and may have loops and multiple edges.} 
$G=(V,E)$ whose edge set
is partitioned into a {\em red edge set\,} $R$ and a {\em blue edge
set\,} $B$. Each red edge $e \in R$ has a nonnegative fixed {\em cost\,}
$c(e)$ (the best competitor's price).
The leader owns every blue edge $e \in B$ and has to set a
{\em price\,} $p(e)$ for each of these edges. The cost function $c$ and
price function $p$ together define a {\em weight\,} function $w$ on
the whole edge set. By ``weight of edge $e$'' we mean either
``cost of edge $e$'' if $e$ is red or ``price of edge $e$'' if
$e$ is blue. A spanning tree $T$ is a {\em minimum
spanning tree\,} (MST) if its {\em total weight}
\begin{equation}
\sum_{e \in E(T)} w(e) = \sum_{e \in E(T) \cap R} c(e)
+ \sum_{e \in E(T) \cap B} p(e)
\end{equation}
is minimum. The {\em revenue\,} of $T$ is then
\begin{equation}
\sum_{e \in E(T) \cap B} p(e).
\end{equation}
The Stackelberg Minimum Spanning Tree problem, \StackMSTnospace, asks for a
price function $p$ that maximizes the revenue of an MST.
Throughout, we assume that the graph contains a spanning tree
whose edges are all red; otherwise, there is a cut consisting only of
blue edges and the optimum value is unbounded. Moreover, to
avoid being distracted by epsilons, we assume that among all edges
of the same weight, blue edges are always preferred to red edges;
this is a standard assumption. As a consequence, all minimum spanning trees
for a given price function $p$ have the same revenue;
see Section \ref{prelim} for details.

\paragraph{Related work.}

A similar pricing problem, where one wants to price the edges in $B$ and
the customer wants to construct a shortest path between two vertices instead 
of a spanning tree, has been studied in the literature; see van Hoesel \cite{vH06} for a
survey. Complexity and approximability results have recently been obtained
by Roch, Savard and Marcotte~\cite{RSM05}, and by 
Bouhtou, Grigoriev,  van Hoesel, van der Kraaij, Spieksma, 
and Uetz~\cite{BGvHvdKSU07}: the problem is strongly NP-hard
and $O(\log |B|)$-approximable. A generalization of the problem
to more than one customer has been tackled using mathematical
programming tools, in particular bilevel programming;
see Labb\'e, Marcotte, and Savard \cite{LMS98}.
This generalization was motivated by the problem of setting tolls on
highway networks. Note that the \StackMST problem is only 
interesting in the single-customer case, since otherwise all customers
purchase the same tree. Cardinal, Labb\'e, Langerman, and Palop~\cite{CLLP05}
give a geometric version of the shortest path problem.

Recently, part of the results of the current paper have been generalized to other problems by Briest, 
Hoefer and Krysta~\cite{BHK08}. They also exhibit a polynomial-time algorithm for a special case of a 
Stackelberg vertex cover problem, in which the follower's problem is to find a minimum vertex cover in a
bipartite graph.

Other pricing problems have been studied, in which 
the goal is to find the best prices for a set of items, after bidders have announced their
preferences in the form of subset valuations. {\em Envy-free} pricing, in particular, 
can be viewed as a simple Stackelberg game.
APX-hardness and approximability of such problems have been established 
by Hartline and Koltum~\cite{HK05}, and by Guruswami, Hartline, Karlin, Kempe, Kenyon, 
and McSherry~\cite{GHKKKS05}. Balcan and Blum~\cite{BB06} gave improved approximation results.
Approximability within a logarithmic factor has also been recently established for more general cases by 
Balcan, Blum and Mansour~\cite{BBM08}. The case in which items are edges of a graph has been 
studied by Grigoriev, van Loon, Sitters and Uetz~\cite{GLSU06}, and Briest and Krysta~\cite{BK06}.
A semi-logarithmic inapproximability result for a special case of the unlimited supply pricing problem
has been given by Demaine, Feige, Hajiaghayi, and Salavatipour~\cite{DFHStoappear}.

\paragraph{Our results.}

We analyze the complexity and approximability of the
\StackMST problem.  Specifically, we prove the following:
\begin{enumerate}
\item \StackMST is APX-hard, even if there are only two red costs,
  $1$ and~$2$ (Section~\ref{Hard}).
  This result is also the first NP-hardness proof for this problem, and, to our knowledge, the first APX-hardness proof 
  for a Stackelberg pricing game with a single customer. The reduction is from \scovernospace.
\item \StackMST is $O(\log n)$-approximable, and is
  $O(1)$-approximable when the red costs either fall in a
  constant-size range or have a constant number of distinct
  values (Section~\ref{sec-best}). More precisely, we analyze
  the following simple approximation algorithm, called
  \emph{Best-out-of-$k$}: for all $i$ between $1$ and~$k$,
  consider the price function for which all blue edges have
  price~$c_i$, and output the best of these $k$ price functions.
  Here, and throughout the paper, $c_i$ denotes the $i$th
  smallest cost of a red edge and $k$ the number of distinct red
  costs. We prove that the approximation ratio of this algorithm
  is bounded above by $\min \{k , 1+\ln b , 1+\ln (c_k/c_1)\}$,
  where $b := |B|$ is the number of blue edges.
\item The integrality gap of a natural integer linear programming formulation
asymptotically matches the approximation guarantee of
  Best-out-of-$k$ (Section~\ref{LP}). Thus, effectively, any
  approximation algorithm based on the linear programming
  relaxation of our integer program (or any weaker relaxation)
  cannot do better than Best-out-of-$k$. Of course, this result
  does not imply that Best-out-of-$k$ is optimal. In fact, a
  central open question about \StackMST is to determine if it
  admits a constant factor approximation algorithm.
\end{enumerate}

\section{Basic Results}
\label{prelim}

Before we proceed to our main results, we prove a few basic lemmas
about \StackMSTnospace.

We claimed in the introduction that the revenue of the leader depends
on the price function $p$ only, and not on the particular MST picked by
the follower. To see this, let $w_1 < w_2 < \cdots < w_\ell$ denote
the different edge weights. The greedy algorithm (a.k.a.\ Kruskal's
algorithm) will work in $\ell$ phases: in its $i$th phase, it will
consider all blue edges of price $w_i$ (if any) and then all red edges
of cost $w_i$ (if any). The number of blue edges selected in the $i$th
phase will not depend on the order in which blue or red edges of weight
$w_i$ are considered. This shows the claim. Moreover, if there is no red
edge of cost $w_i$ then $p$ is not an optimal price function because the
leader can raise the price of every blue edge of price $w_i$ to the next
weight $w_{i+1}$ and thus increase his/her revenue. This implies
the following lemma.

\begin{lemma}
\label{AssignedCostEqualGivenCost}
In every optimal price function, the
prices assigned to the blue edges appearing in some MST
belong to the set $\{c(e) : e \in R\}$. 
\end{lemma}

Notice that for optimal price functions, the prices given to the blue 
edges that are in no MST do not really matter, as long as they are high enough. 
We find it convenient to see them as equaling $\infty$. This has the same
effect as deleting those blue edges. A direct consequence of
Lemma~\ref{AssignedCostEqualGivenCost} is that the decision version
of \StackMST belongs to NP, using some price function $p$ with $p(e)
\in \{c(e) : e \in R\} \cup \{\infty\}$ for all $e \in B$
as a certificate. Another possibility for a certificate is
an acyclic set of blue edges $F$, interpreted as
the set of blue edges in any MST. Given $F$, we can easily compute an
optimal price function such that $F$ is the set of blue edges in any
MST, with the help of Lemma \ref{lemma-cuts} below. In the lemma, we
use the notation $\mathcal{C}(B',e)$ for the set of cycles of $G=(V,R \cup B')$ that
include the edge $e$, where $B'$ is an acyclic subset of blue edges and $e\in B'$.
(Notice that $\mathcal{C}(B',e)$ is nonempty because $(V,R)$ is connected.)

\begin{lemma}
\label{lemma-cuts} Consider a price function $p$, a corresponding
minimum spanning tree $T$, and let $F= E(T) \cap B$. Then for every
$e\in F$, we have
\begin{equation}
\label{eq:min-max}
p(e) \le \min_{C \in \mathcal{C}(F,e)} \max_{e'\in E(C) \cap R}c(e').
\end{equation}
Moreover, whenever $F$ is any acyclic set of blue edges and we set
$p(e)$ equal to the right hand side of \eqref{eq:min-max} for $e \in F$
and $p(e) = \infty$ for $e \in B - F$, we have $E(T') \cap B = F$ for
any corresponding MST~$T'$.
\end{lemma}
\begin{proof}
The first part of the lemma is straightforward. Indeed, if \eqref{eq:min-max}
fails for some edge $e \in F$, then there exists a red edge $e'$
with $c(e') < p(e)$ that links the two components of $T - e$, and so $T$
cannot be an MST. We now turn to the second part of the lemma. First note
that $E(T') \cap B$ is clearly contained in $F$ because no MST can use
any edge with an infinite price. By contradiction, suppose there is some
edge $e$ in $F$ that is not used by $T'$ and let $e'$ be a red edge with
maximum cost on the unique cycle of $T' + e$. Because the price function $p$
we have chosen satisfies \eqref{eq:min-max} (with equality), the weight
of edge $e$ is at most the weight of $e'$, and thus $T'$ is not an MST
because of our assumption that blue edges have priority over the red edges
of the same weight. 
\end{proof}

It follows from the above lemma that \StackMST is fixed parameter
tractable with respect to the number of blue edges. Indeed, to solve
the problem, one could try all acyclic subsets $F$ of $B$, and for
each of them put the prices as above (this can easily be done in
polynomial time), and finally take the solution yielding the highest
revenue.
We conclude this section by stating a useful property satisfied
by all optimal solutions of \StackMSTnospace.

\begin{lemma}
\label{obstruction}
Let $p$ be an optimal price function and $T$ be a corresponding MST.
Suppose that there exists a red edge $e$ in $T$ and a blue edge $f$
not in $T$ such that $e$ belongs to the unique cycle $C$ in $T + f$.
Then there exists a blue edge $f'$ distinct from $f$ in $C$ such that
$c(e) < p(f') \le p(f)$.
\end{lemma}
\begin{proof}
The inequality $c(e) < p(f)$ follows from the optimality of $T$ and
from our assumption on the priority of blue edges versus red edges
of the same weight. If all blue edges $f'$ distinct from $f$ in $C$
satisfied $p(f') \le c(e)$ or $p(f) < p(f')$ then by decreasing
the price of $f$ by some amount we would be able to find a new
price function $p'$ such that $T' = T-e'+f$ is an MST with respect to
$p'$, where $e'$ is some red edge on $C$. This contradicts the
optimality of $p$ because the revenue of $T'$ is bigger than that
of $T$. 
\end{proof}

\section{Complexity and Inapproximability}
\label{Hard}

By Lemma~\ref{AssignedCostEqualGivenCost}, \StackMST is trivially
solved when the cost of every red edge is exactly $1$, i.e.,
when $c(e) = 1$ for all $e\in R$.
In this section, we show that the problem is APX-hard
even when the costs of the red edges are only $1$ and $2$,
i.e., when $c(e)\in \{1,2\}$ for all $e\in R$.
We start with NP-hardness:

\begin{theorem}
\label{Hardness} \StackMST is NP-hard even when $c(e)\in \{1,2\}$
for all $e\in R$.
\end{theorem}
\begin{proof}
We present a reduction from \scover (in its decision version).
Let $(\mathcal{U,S})$ and the integer $t$ be an instance
of \scovernospace, where $\mathcal{U}=\{u_1,u_2,\ldots,u_n\}$,
and $\mathcal{S} = \{S_1,S_2,\ldots,S_m\}$. Without loss of
generality, we assume that $u_n \in S_i$ for every $i=1,2,\ldots,m$
(we can always add one element to $\mathcal{U}$ and to every $S_i$
to make sure this holds).

We construct a graph $G=(V,E)$ with edge set $E= R \cup B$ and a cost
function $c : R \rightarrow \{1,2\}$ as follows. The vertex set of
$G$ is $\mathcal{U} \cup \mathcal{S} = \{u_1,u_2,\ldots,u_n\} \cup
\{S_1,S_2,\ldots,S_m\}$. The edge set of $G$ and cost function $c$
are defined as follows:
\begin{itemize}
\item there is a red edge of cost $1$ linking $u_i$ and $u_{i+1}$
      for every $1 \leq i < n$;
\item there is a red edge of cost $2$ linking $u_n$ and $S_1$, and
      linking $S_j$ and $S_{j+1}$ for every $1 \leq j < m$;
\item whenever $u_i \in S_j$ we link $u_i$ and $S_j$ by a blue edge.
\end{itemize}
\begin{figure}[h!]
\begin{center}
\includegraphics[scale=0.48]{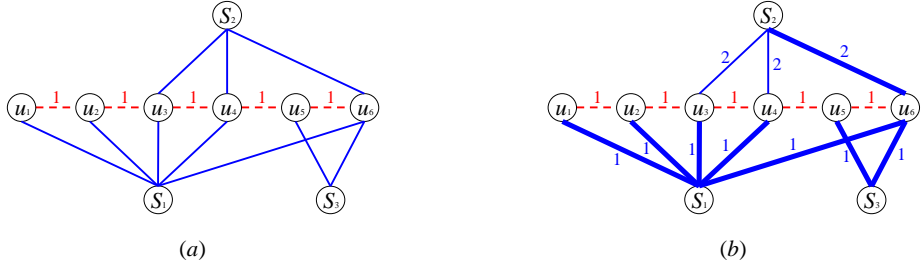}
\caption{\label{Figure:Construction}(a) The graph $G$ constructed
for $n=6$, $m=3$ 
with $S_1=\{u_1,u_2,u_3,u_4,u_6\}$,
$S_2=\{u_3,u_4,u_6\}$ and $S_3=\{u_5,u_6\}$. The red edges of
cost $2$ are omitted for clarity. The red edges of cost $1$
are dashed, and the blue edges are solid. (b) An optimal price
function $p$ on the blue edges that yields a revenue of $9$, an
example MST is depicted in bold. 
}
\end{center}
\end{figure}
We illustrate such a construction in Fig.~\ref{Figure:Construction}.
We claim that $(\mathcal{U,S})$ has a set cover of size $t$ if and only if
there exists a price function $p : B \rightarrow \{1,2,\infty\}$ for
the blue edges of $G$ whose revenue is $n+2m-t-1$.

\medskip

\noindent ($\Rightarrow$) Suppose $(\mathcal{U,S})$ has a set cover of size
$t$. We construct $p$ as follows: for every blue edge $e=u_iS_j$,
we set $p(e)$ to be $1$ if $S_j$ is in the set cover, and $2$
otherwise. We show that the revenue of $p$ equals $n+2m-t-1$ by
running Kruskal's MST algorithm starting with an empty tree, $T$.
Because the blue edges of weight $1$ are the lightest, we start with
adding them one by one to $T$ such that we add an edge only if it
doesn't close a cycle in $T$. After going over all blue edges of
weight $1$, we are guaranteed that $T$ is a tree that spans all the
vertices $u_i$ for every $i = 1,\ldots ,n$, and every vertex $S_j$
such that $S_j$ is in the set cover. This is because these vertices are
connected through $u_n$ with only blue edges of weight $1$. So the
current weight of $T$ is $|T|-1=n+t-1$. We next try to add the red
edges of weight $1$, but every such edge connects two vertices,
$u_i$ and $u_{i+1}$, already spanned by $T$ and therefore closes
a cycle, so we add none of them. Next we add the blue edges of weight
$2$. For every $S_j$ not in the set cover, we connect $S_j$ to $T$
with one blue edge of weight $2$ (the second one will close a cycle).
Therefore, after going over all the blue edges of weight $2$, we
added a weight of $2(m-t)$ to $T$. Furthermore, $T$ spans the entire
graph so there is no need to add any red edges of weight $2$. All the
edges in $T$ are blue and the revenue of $T$ is $(n+t-1)+2(m-t)=n+2m-t-1$.

\medskip

\noindent ($\Leftarrow$) Suppose that there exists a price function $p : B
\rightarrow \{1,2,\infty\}$ for the blue edges of $G$ whose revenue is
$n+2m-t-1$ for some $t$. By Lemma~\ref{AssignedCostEqualGivenCost},
there exists such a function $p$ that is optimal. Choose then $p : B
\rightarrow \{1,2,\infty\}$ as an optimal price function that minimizes
the number of red edges in an MST $T$.

Assume first that $T$ contains only blue edges. Then every vertex $u_{i}$ is
incident to some blue edge in $T$ with price 1. To see this, observe that $u_i$ is
adjacent to a vertex $S_j$ that is not a leaf, thus $S_j$ has a neighbor $u_k$, and
the red edges in the cycle $S_j,u_1,\ldots ,u_k,S_j$ all have cost 1.
Thus the set $\mathcal{S}'$ of those $S_{j}$'s that are linked to some blue 
edge in $T$ with price 1 is a set cover of $(\mathcal{U,S})$. On the other hand, notice that
any $S_j \in \mathcal{S} \setminus \mathcal{S}'$ is a leaf of $T$, because
if there were two blue edges $u_{i}S_{j}, u_{i+\ell}S_{j}$ in $T$ then none of
them could have a price of 2 because of the cycle $S_{j}u_{i}u_{i+1}\dots u_{i+\ell}S_j$. 
Therefore, the revenue of $p$ equals $(n + |\mathcal{S}'| - 1) 
+ 2(m - |\mathcal{S}'|) =  n + 2m - |\mathcal{S}'| - 1$. As by hypothesis
this is at least $n + 2m - t - 1$, we deduce that the set cover $\mathcal{S}'$
has size at most $t$.

Suppose now that $T$ contains some red edge $e$ and denote by $X_1$
and $X_2$ the two components of $T-e$. There exists some blue edge
$f=u_i S_j$ in $G$ that connects $X_1$ and $X_2$ because
the graph $(V,B)$ induced by the blue edges is connected (because $u_n$
is linked with blue edges to every $S_j$). By Lemma \ref{obstruction},
there exists a blue edge $f'=u_{i'}S_{j'}$ distinct from $f$ in the unique
cycle $C$ in $T+f$ such that $c(e) < p(f') \le p(f)$. In particular, we
have $c(e) = 1$ and $p(f') = 2$. By an argument given in the preceding
paragraph, $S_{j'}$ is a leaf of $T$, hence we have $j' = j$. Also,
every blue edge distinct from $f$ and $f'$ in $C$ has price $1$.
But then the price function $p'$ obtained from $p$ by setting the price of
both $f$ and $f'$ to 1 is also optimal and has a corresponding MST that uses
less red edges than $T$, namely $T-e+f$, a contradiction. This completes
the proof. 
\end{proof}
The reduction used in Theorem~\ref{Hardness} implies a stronger
hardness result.
\begin{theorem}
\label{APX_Hardness} \StackMST is APX-hard even when $c(e) \in \{1,2\}$
for all $e\in R$.
\end{theorem}
\begin{proof}
We will show that, for any $\varepsilon > 0$, a $(1-\varepsilon)$-approximation
for \StackMST implies a $(1+8 \varepsilon)$-approximation for
\vcover in graphs of maximum degree at most $3$. The claim will then follows
from the APX-hardness of the latter problem~\cite{VertexCoverOnBoundedDegree3,VertexCoverOnBoundedDegree}.

Let $H$ denote any given graph with maximum degree at most $3$.
We can assume that $H$ is connected because otherwise we process each
connected component separately. Moreover, we can assume that $H$ has at
least as many edges as vertices because \vcover can be solved exactly in
polynomial time if $H$ is a tree.

Clearly, the \vcover instance we consider is equivalent to a \scover
instance with $|V(H)|$ sets and $|E(H)|$ elements in the ground set.
Let $(\mathcal{U,S})$ be the \scover instance obtained from the latter
one by adding a new dummy element $d$ in the ground set, and adding $d$
to every subset of the instance. Hence, we have $n = |\mathcal{U}| =
|E(H)| + 1$ and $m = |\mathcal{S}| = |V(H)|$. Any vertex cover of $H$
yields a set cover of $(\mathcal{U,S})$ with the same size, and vice-versa.
Thus the reduction used in the proof of Theorem~\ref{Hardness} provides a
way to convert in polynomial time a vertex cover of size $s$ into a feasible
solution of the \StackMST instance corresponding to $(\mathcal{U,S})$
with revenue $n + 2m - s - 1$, and vice-versa. In particular, we have
$\OPT = n + 2m - \OPTVC - 1$, where $\OPT$ and $\OPTVC$ denote the
value of the optimum for the \StackMST and \vcover instances, respectively.

Now consider the vertex cover found by running the
$(1-\varepsilon)$-approximation algorithm on the \StackMST instance
and then converting the result into a vertex cover of $H$.
Denoting by $s$ its size and letting $r = n+2m-s-1$, we obtain:
$$
\begin{array}{r@{\quad}c@{\quad}l}
s = n + 2m - r-1 &\le& n + 2m - (1-\varepsilon) \, \OPT - 1\\[1ex]
&=& n + 2m - (1-\varepsilon) \, (n + 2m - \OPTVC - 1) - 1\\[1ex]
&=& \varepsilon \, (n - 1 + 2m) + (1-\varepsilon)  \, \OPTVC \\[1ex]
&\le& \varepsilon \, (3 \, \OPTVC + 6 \, \OPTVC) + (1-\varepsilon) \, \OPTVC \\[1ex]
&=& (1 + 8 \varepsilon)  \, \OPTVC.
\end{array}
$$
Above we have used the fact that $n - 1 = |E(H)| \ge |V(H)| = m$ and
that $\OPTVC \ge |E(H)|/3 = (n-1)/3$ because $H$ has maximum degree at
most~$3$.
\end{proof}

\section{The Best-Out-Of-$k$ Algorithm}
\label{sec-best}

As before, let $k$ denote the number of distinct red costs, and
let $c_1 < c_2 < \cdots < c_k$ denote those costs. Without loss
of generality, we assume that all red costs are positive (otherwise
we contract all red edges of cost $0$). Recall that the Best-out-of-$k$
algorithm is as follows. For each $i$ between $1$ and $k$, set $p(e) =
c_i$ for all blue edges $e \in B$ and compute an MST $T_i$. Then pick
$i$ such that the revenue of $T_i$ is maximum and output the
corresponding feasible solution. In this section, we analyze the approximation 
ratio ensured by this algorithm. 

\begin{theorem}\label{best-out-of-k}
Best-out-of-$k$ is a
$\min \{k , 1+\ln b, 1+\ln W\}$-approximation algorithm,  where $b$ denotes the
number of blue edges, and $W = c_k / c_1$ is the maximum ratio between red costs.
\end{theorem}
\begin{proof}
We let $p^*$ be an optimal price function, $T^*$ be an MST of $G$ with respect to $p^*$, and $n_i$ be the number of blue edges of price $c_i$ in $T^*$. We also define $N_i$ as the number of blue edges of price at least $c_i$ in $T^*$, that is, $N_i = \sum_{j=i}^k n_j$.

We first prove the following claim: for all $i = 1, \ldots, k$, the $i$th MST $T_i$ computed by Best-out-of-$k$ contains at least $N_i$ blue edges. 
For $S \subseteq E$, let $r(S)$ denote the maximum cardinality of an acyclic subset of $S$ (that is, the rank function of the graphic matroid of $G$).
We also let $R_{i}$ be the set of red edges with cost at most $c_i$, and $B^*_{i}$ be the set of blue edges $e$ such that $p^*(e) \le c_i$. 

Now consider an execution of Kruskal's algorithm on $G$ with respect to $p^*$, up to the point where all edges of weight at most $c_{i-1}$ have been processed. The total number of edges included up to that point in the MST $T^*$ equals $r(R_{i-1}\cup B^*_{i-1})$. Next, resume the execution of Kruskal's algorithm, process all blue edges of price $c_i$ and stop before processing any red edge of cost $c_i$. In order to maximize the number of blue edges $N_i$ of price at least $c_i$ included in $T^*$, we could lower to $c_i$ the price of all blue edges whose current price is at least $c_i$. Then, the total number of edges included up to now in $T^*$ would be exactly $r(R_{i-1}\cup B)$. This implies
$$
N_i \leq r(R_{i-1}\cup B) - r(R_{i-1}\cup B^*_{i-1}) \leq  r(R_{i-1}\cup B) - r(R_{i-1}).
$$
Because the latter expression gives the number of blue edges in $T_i$, this proves the claim.

Using this claim, we can bound the revenue $q$ of the solution returned by Best-out-of-$k$:
$$
q \geq \max_{i=1,\ldots,k} N_i\cdot c_i.
$$
We also know that $OPT = \sum_{i=1}^k n_i\cdot c_i$.

Since $n_i\leq N_i$, we have 
$$
OPT = \sum_{i=1}^k n_i\cdot c_i\leq \sum_{i=1}^k N_i\cdot c_i\leq k\cdot q,$$ 
proving the first approximation factor.

Also, we have (letting $N_{k+1}=0$):
\begin{eqnarray*}
OPT & = & \sum_{i=1}^k n_i\cdot c_i \\
	& = & \sum_{i=1}^k N_i \cdot c_i \cdot \frac{n_i}{N_i} \\
	& = & \sum_{i=1}^k N_i \cdot c_i \cdot \frac{N_i - N_{i+1}}{N_i} \\
	& \leq & (\max_{i=1,\ldots,k} N_i \cdot c_i) \cdot \sum_{i=1}^k  \frac{N_i - N_{i+1}}{N_i} \\
	& \leq & q \cdot \sum_{i=1}^k  \frac{N_i - N_{i+1}}{N_i}, 
\end{eqnarray*}
and
$$
\sum_{i=1}^k  \frac{N_i - N_{i+1}}{N_i} \leq 1+\int_{N_k}^{N_1} \frac{dt}{t} \leq 1+\ln\frac{N_1}{N_k}\leq 1+\ln b,
$$
which proves the second approximation  factor.

Finally, we also have the following (letting $c_0=0$):
\begin{eqnarray*}
OPT & = & \sum_{i=1}^k n_i\cdot c_i \\
	& = & \sum_{i=1}^k n_i \sum_{j=1}^i (c_j - c_{j-1}) \\
	& = & \sum_{j=1}^k N_j\cdot (c_j - c_{j-1}) \\
	&\leq & q\cdot \sum_{j=1}^k \frac{c_j - c_{j-1}}{c_j},
\end{eqnarray*}
and
$$
\sum_{j=1}^k \frac{c_j - c_{j-1}}{c_j} \leq 1+\ln W,
$$
establishing the third approximation factor.
\end{proof}

The three approximation factors are tight for the following examples. Consider
a graph with $k+1$ vertices $v_1,v_2,\ldots ,v_{k+1}$, in which the red edges 
are of the form $v_iv_{i+1}$, and there is a blue edge parallel to every red edge.
The cost of the red edge $v_iv_{i+1}$ is $1/i$. The optimal solution involves setting a
price of $1/i$ for every blue edge $v_iv_{i+1}$, yielding a revenue of $\sum_{i=1}^k 1/i$.
On the other hand, the Best-out-of-$k$ algorithm sets the price of every blue edge to $1/i$
for some $i$, always yielding a revenue of 1. This proves that the ratios $1+\ln b$ and 
$1+\ln W$ are asymptotically tight.

The factor $k$ can be proven tight as well by considering a similar example.
The graph is composed of $1+\sum_{i=1}^k a^{i-1}$ vertices for some large integer $a$.
The red edges form a path connecting these vertices using $a^{k-i}$ edges of cost
$c_i=a^{i-1}$ for every $i$ between 1 and $k$. Every red edge is doubled by a blue edge.
The optimal solution again involves setting the prices of the blue edges equal to that of the 
parallel red edge, yielding a revenue of $k\cdot a^{k-1}$. The Best-out-of-$k$ algorithm
setting the prices to $c_i$ yields an MST containing $\sum_{j=i}^k a^{k-j}$ blue edges, with a revenue
of
$$
a^{i-1}\cdot \sum_{j=i}^k a^{k-j} = a^{i-1}\cdot \frac{a^{k-i+1} - 1}{a-1} 
< a^{k-1}\cdot \frac{a}{a-1}.
$$
The ratio between the two revenues tend to $k$ as $a$ tends to infinity.\medskip

A natural generalization of \StackMST to matroids is
as follows. Given a matroid $(S, \mathcal{I})$ with $\mathcal{I}$
partitioned into two sets $\mathcal{R}$ and $\mathcal{B}$, and
nonnegative costs on the elements of $\mathcal{R}$, assign prices on
the elements of $\mathcal{B}$ in such a way that the revenue given
by a minimum weight basis of $(S, \mathcal{I})$ is maximized. We mention
 that the analysis of Best-out-of-$k$
given in the proof of Theorem~\ref{best-out-of-k} extends swiftly
to the case of matroids, yielding the same approximation for
this more general case.

\section{Linear Programming Relaxation}
\label{LP} In this section, we give an integer programming
formulation for the problem and study its linear
programming relaxation. All red costs $c_{i}$ are assumed to be positive
throughout the section. 
For each $j=1, \ldots, k$, and
each blue edge $e \in B$ we define a variable $x_{j,e}$. The
interpretation of these variables is as follows: think of a feasible
solution $p : B \to \{c_1,c_2,\ldots,c_k\}$ and an MST $T$
with respect to $p$.
Then $x_{j,e} = 1$ means that the blue edge $e$ appears
in $T$, with a price $p(e)$ of at least $c_j$.

We let $c_0 = 0$ and, as in the previous section, denote by $R_j$ the set of red
edges of cost at most $c_j$.
For $t$ pairwise disjoint sets of vertices $C_1, \dots, C_t$, we denote by
$\delta_B(C_1:C_2:\cdots:C_t)$ the set of blue edges that are in the
cut defined by these sets.
The integer programming
formulation then reads:
\begin{align}
\nonumber
(\IP) \quad \textrm{max}
&\ds \sum_{e \in B \atop 1 \le j \le k} (c_j - c_{j-1})\,x_{j,e}\\[4ex]
\textrm{s.t.}
&\ds \sum_{e \in \delta_B(C_1:C_2:\cdots:C_t)} \!\!\!\!\!\!\!\!\!\!\! \!\! x_{j,e} \le t - 1
  & &\forall j \in \{1,2,\dots,k\},\label{IP-forest}\\[-2ex]
\nonumber & & &\forall C_1, ..., C_t \textrm{ components of } (V,R_{j-1});\\[1ex]
&\ds \sum_{e \in P \cap B} x_{1,e} + x_{j,f} \le |P \cap B|
  & &\forall f=ab \in B, \forall j \in \{2,3,\dots,k\}, \label{IP-cycle}\\[-2ex]
\nonumber & & &\forall P \textrm{ $ab$-path in } (B \cup R_{j-1}) - f;\\[1ex]
&\ds x_{1,e} \ge x_{2,e} \ge \cdots \ge x_{k,e} \ge 0
  & &\forall e \in B; \label{IP-equality} \\[1ex]
&x_{j,e} \in \{0,1\} & & \forall j \in \{1,2,\dots,k\}, \forall e \in B. \label{IP-integer}
\end{align}

Let us first give some intuition on this integer program.
Consider a minimum spanning tree $T$ with respect to a feasible solution $p$, let $F$ be the set of blue edges appearing in $T$, and let $F_{j} =\{e\in F: p(e) \geq c_{j}\}$.
Then $F$ ($=F_{1}$) must obviously be a forest. 
Also, $F_{j}$ ($j \in \{2,\dots,k\}$) must be a forest in the
graph where every component of $(V, R_{j-1})$ has been contracted, since otherwise
we could swap in $T$ some edge of $F_{j}$ with an edge in $R_{j-1}$.
This is encoded by constraints~\eqref{IP-forest}.
Similarly, if a cycle $C$ of the graph is such that every red edge in $C$
has cost at most $c_{j-1}$ and some blue edge $f$ of $C$ appears in $T$ with a
price at least $c_{j}$, then there must be another blue edge of $C$ that is not included in $T$.
This is ensured by constraints~\eqref{IP-cycle}. 

\begin{proposition}
The integer program above is a formulation of \StackMSTnospace.
\end{proposition}
\begin{proof}
Consider a feasible solution $x$ of the integer program (IP) and
let $F = \{e \in B : x_{1,e} = 1\}$. Inequality~\eqref{IP-forest} ensures
that $F$ is a forest. For $e \in F$, let $p(e) = c_j$ if $j$ is
the last index for which $x_{j,e} = 1$ and, for $e \in B-F$,
let $p(e) = \infty$. Now consider a minimum spanning tree $T$
with respect to $p$. We claim $E(T) \cap B = F$ and that the revenue of $T$ is exactly the
objective value for $x$.

It suffices to prove that all edges of $F$ belong to $T$. All
edges $e \in F$ of price $c_1$ are necessarily in $T$. Assume
that all edges $e \in F$ of price less than $c_j$ are in $T$, for
some $j \ge 2$. We show that this holds too for edges of price
$c_j$. Consider some edge $f$ with $p(f) = c_j$. Suppose that $f$
is not in $T$. This means that there exists a cycle in $G$
consisting of blue edges of price at most $c_j$ and of red
edges of price at most $c_{j-1}$. But then~\eqref{IP-cycle} is violated,
a contradiction. So the claim holds.

Conversely, consider any optimal solution to the \StackMST problem
with price function $p(\cdot)$ and a corresponding MST $T$.
Let $F = E(T) \cap B$. We define a vector $x$ as follows: for $e\in B$,
$x_{i,e}=1$ if $e\in F$ and $p(e) \ge c_i$, otherwise $x_{i,e}=0$. It is
easily checked that the revenue given by $p$ equals the objective function
of the IP for $x$. Moreover, constraints~\eqref{IP-forest}, \eqref{IP-equality}
and~\eqref{IP-integer} are clearly satisfied by $x$. Finally, note that if
$x$ violates~\eqref{IP-cycle} for some $e\in F$, then $e$ also violates
the min-max formula given in Lemma~\ref{lemma-cuts}. This completes the proof.
\end{proof}

The rest of this section is devoted
to the LP relaxation of the above IP, obtained by dropping constraint
\eqref{IP-integer}. We show that the LP is tractable and that its integrality gap 
matches essentially the guarantee given by the Best-out-of-$k$ algorithm.
(Let us recall that the integrality gap of the LP on a specified set of instances $\mathcal{I}$
is defined as the supremum of the ratio $(\LP) / (\IP)$ over all instances in $\mathcal{I}$.)

\begin{proposition}
\label{prop-separation}
The LP can be separated in polynomial time.
\end{proposition}
\begin{proof}
For fixed $j$, \eqref{IP-forest} can be separated in polynomial time using
standard techniques for the forest polytope, as described e.g. in Schrijver~\cite[pp. 880--881]{S03B}.
Inequality \eqref{IP-cycle} can be rewritten as
$$
\sum_{e \in P \cap B} (1 - x_{1,e}) \ge x_{j,f}.
$$
Thus, for each fixed $j$ and $f=ab$, \eqref{IP-cycle} can be separated by finding
a shortest $ab$-path in the graph $(V, (B \cup R_{j-1}) - f)$ where every
red edge has weight 0 and every blue edge $e$ has weight $1-x_{1,e}$.
Finally,  \eqref{IP-equality} can obviously be separated in polynomial time.
\end{proof}

We first bound the integrality gap from above:

\begin{proposition}
\label{prop-LP-bound}
We have $(\LP) \le \min \{k , 1+\ln b, 1+\ln W\} \cdot (\IP)$,  where $b$ denotes the
number of blue edges, and $W = c_k / c_1$ is the maximum ratio between red costs.
\end{proposition}
\begin{proof}
Let $x$ be any feasible vector for the LP. The value
of the objective function for $x$ is thus 
$$
\sum_{e \in B \atop 1 \le i \le k} (c_i - c_{i-1})\,x_{i,e}.
$$

Let $i\in \{1,\dots,k\}$, 
let $C^{1}, \dots, C^{\ell}$ be components of the graph $(V, R_{i-1} \cup B)$,
and denote by $C^{j}_{1}, \dots, C^{j}_{\ell_{j}}$ the components of the subgraph of
$(V, R_{i-1})$ induced by $C^{j}$. For every $j\in \{1,\dots, \ell\}$, we have
$$
\sum_{e \in B[C^{j}_{1} \cup \cdots \cup C^{j}_{\ell_{j}}]} x_{i,e}
= \sum_{e \in \delta_B(C^{j}_{1}:C^{j}_{2}:\cdots:C^{j}_{\ell_{j}})} x_{i,e}.
$$
(Here, for $S\subseteq V$, the notation $B[S]$ means the set of blue edges with both endpoints
in $S$.)
Indeed, this holds trivially if $i=1$, since then each $C^{j}_{p}$ is a vertex of $C^{j}$.
For $i\geq 2$, for any blue edge $f=ab$ that is internal to a component
$C^{j}_{p}$ of $C^{j}$ (that is, $f \in B[C^{j}_{p}]$), there exists an $ab$-path
consisting of edges of $R_{i-1}$, and so~\eqref{IP-cycle} enforces that $x_{i,f} \leq 0$.

Also, constraints~\eqref{IP-forest} imply 
$$
\sum_{e \in \delta_B(C^{j}_{1}:C^{j}_{2}:\cdots:C^{j}_{\ell_{j}})} x_{i,e} \le \ell_{j} - 1,
$$
for every $j\in \{1,\dots, \ell\}$.
We thus obtain
$$
\sum_{e \in B} x_{i,e} 
= \sum_{j=1}^{\ell} \sum_{e \in \delta_B(C^{j}_{1}:C^{j}_{2}:\cdots:C^{j}_{\ell_{j}})} x_{i,e}
 \le \sum_{j=1}^{\ell} (\ell_{j} - 1) = r(R_{i-1}\cup B) - r(R_{i-1}).
$$
The number of blue edges in the $i$th MST computed by  Best-out-of-$k$ being exactly
 $r(R_{i-1}\cup B) - r(R_{i-1}) =: A_{i}$, it then follows
$$
\sum_{e \in B \atop 1 \le i \le k} (c_i - c_{i-1})\,x_{i,e} \leq \sum_{i=1}^{k} (c_i - c_{i-1})\,A_{i}.
$$
Letting $q=\max_{i=1,\ldots,k} A_{i}\cdot c_{i}$
 denote the revenue given by the Best-out-of-$k$ algorithm, we deduce
$$
 \sum_{i=1}^{k} (c_i - c_{i-1}) A_{i}
 =  \sum_{i=1}^{k} \frac{c_i - c_{i-1}}{c_{i}} A_{i}\cdot c_{i}
 \leq q\cdot \sum_{i=1}^{k} \frac{c_i - c_{i-1}}{c_{i}},
$$
and, letting $A_{k+1}=0$,
$$
 \sum_{i=1}^{k} (c_i - c_{i-1}) A_{i}
 =  \sum_{i=1}^{k} c_i (A_{i} - A_{i+1})
=  \sum_{i=1}^{k} A_{i}\cdot c_i \frac{A_{i} - A_{i+1}}{A_{i}}
\leq q\cdot \sum_{i=1}^{k} \frac{A_{i} - A_{i+1}}{A_{i}}. 
$$
As in the proof of Theorem~\ref{best-out-of-k}, we have 
$$
\sum_{i=1}^{k} \frac{c_i - c_{i-1}}{c_{i}} \leq \min\{k, 1 + \ln W\}
$$
and
$$
\sum_{i=1}^{k} \frac{A_{i} - A_{i+1}}{A_{i}} \leq 1 + \ln b.
$$
Therefore, 
\begin{align*}
\sum_{e \in B \atop 1 \le i \le k} (c_i - c_{i-1})\,x_{i,e} 
&\leq \min \{k , 1+\ln b, 1+\ln W\} \cdot q \\
&\leq \min \{k , 1+\ln b, 1+\ln W\} \cdot (\IP),
\end{align*}
as claimed.
\end{proof}

\begin{proposition}
\label{prop-gap}
The integrality gap of the LP is 
\begin{itemize}
\item $k$ on instances with $k$ distinct costs;
\item $\Theta(\ln W)$ on instances with maximum ratio between red costs $W$, and
\item $\Theta(\ln b)$ on instances with $b$ blue edges.
\end{itemize}
\end{proposition}
\begin{proof}
We already know from Proposition~\ref{prop-LP-bound} that
the integrality gap of the LP is at most $\min \{k , 1+\ln b, 1+\ln W\}$.
We first by prove that the integrality gap is at least $k$ on instances with $k$
distinct costs. To this aim, we define an instance of \StackMST as follows:
Choose an integer $a\ge 2$ and let the vertex set of the graph be
$V=\{0,1,2,\dots,a^{k-1}\}$. The graph has $a^{k-1}$ blue edges, linking vertex $0$
to every other vertex. The $i$th red cost is $c_i = a^{i-1}$.
For $i\in \{1,2,\dots,k-1\}$, the subgraph spanned by the
red edges with cost $c_{i}$ is a disjoint union of $a^{k-i-1}$ cliques, each
of cardinality $a^{i}$; the vertex sets of these cliques are
$\{1,\dots,a^{i}\}, \{a^{i} + 1,\dots,2a^{i}\}, \dots, \{a^{k-1} - a^{i} + 1,\dots,a^{k-1}\}$.
Finally, there is a unique red edge with cost $c_{k}$, linking vertex $0$ to vertex $1$.

Consider an optimal solution of the \StackMST problem for the instance defined above,
and let $T$ be a corresponding MST. Consider any blue edge $e$
in $T$, of price $c_i$, and let $C_e$ be the unique component of $(V - \{0\},R_{i-1})$
that contains an endpoint of $e$.
No other blue edge of $T$ has an endpoint in $C_e$, because
otherwise one could replace the edge $e$ in $T$ with an appropriate red edge of
$R_{i-1}$ and obtain a new spanning tree with weight strictly less than that of $T$, a
contradiction.
Thus, if $e$ and $f$ are two distinct blue edges of $T$, then
$C_e \cap C_f = \emptyset$. Noticing that the price given to $e$ is $c_i = a^{i-1} = |C_e|$, we deduce that
the revenue given by $T$ is
$$
\sum_{e\in B\cap E(T)} |C_e| \le a^{k-1}.
$$
Moreover, a revenue of $a^{k-1}$ is easily achieved, set for instance all blue
edges of the graph to the same price $c_i$ for some $i\in \{1,\dots,k\}$. Hence,
$(\IP)=a^{k-1}$.

We now define a feasible solution $x^*$ for the LP. The point $x^*$ will have the property that
$x^*_{i,e}=x^*_{i,f}$ for $1\le i \le k$ and all $e,f \in B$.
We thus let $y_i=x^*_{i,e}$ for $e\in B$. The constraints on the $y_i$'s imposed by the LP are then:
\begin{align*}
&a^{i-1}y_i \le 1 &  \textrm{ for } 1\le i\le k;\\
&y_1 + y_i \le 1 &  \textrm{ for } 2\le i\le k;\\
&y_1 \ge y_2 \ge \cdots \ge y_k \ge 0.
\end{align*}

Set $y_1 = (a-1)/a$ and $y_i = 1/a^{i-1}$ for $2 \le i \le k$, which satisfies
the above constraints. The value of the objective function of the LP for the point $x^*$ is
\begin{align*}
\LP(x^*)&=\sum_{e \in B \atop 1 \le i \le k} (c_i - c_{i-1}) x^*_{i,e}\\
&= a^{k-1}\left(\frac{a-1}{a} + \sum_{2 \le i \le k} (a^{i-1} - a^{i-2}) \frac{1}{a^{i-1}}\right)
= ka^{k-1} - ka^{k-2}.
\end{align*}
Therefore, the ratio $\LP(x^*) / (\IP)$ tends to $k$ as $a \to \infty$.

Now, the same construction can be used to show that the integrality gap
is $\Omega(\ln W)$ and $\Omega(\ln b)$ on instances with $c_{k}/c_{1} = W$
and $b$ blue edges, respectively. We explain it in the case where the number
of blue edges is fixed to some value $b$, the case 
where the ratio $c_{k}/c_{1}$ is fixed is done similarly.

Take an instance as above, with $a=2$ and $k$ being the greatest integer 
such that $2^{k-1} \leq b$. Choose an arbitrary blue edge and add $b - 2^{k-1}$ 
parallel blue edges to it (so that the number of blue edges is exactly $b$). 
These extra blue edges have clearly no influence on the value 
of $(\IP)$ and $\LP(x^*)$ (where $x^*$ is defined as before). Using
$b < 2^{k}$, we deduce
$$
\frac{\LP(x^*)}{(\IP)} = \frac{k2^{k-1} - k2^{k-2}}{2^{k-1}} = \frac{k}{2} > \frac{\log_{2}b}{2},
$$
and thus that the integrality gap is $\Omega(\ln b)$, as claimed.
\end{proof}

To conclude this section, let us mention that
we know of additional families of valid
inequalities that
cut the fractional point used in the above proof.
We leave the study of those for future research.

\section*{Acknowledgments}
We thank Martine Labb\'e and Gilles Savard for preliminary
discussions concerning this problem, Martin Hoefer for his comments 
which led us to refine our approximability result. 
We are also most grateful to the second 
anonymous referee  for providing us with a much shorter proof of Theorem~\ref{best-out-of-k},
and for her or his many insightful remarks which led to an improved version of the paper.

\bibliographystyle{plain}
\bibliography{MST}
\end{document}